\documentclass{article}

\usepackage{arxiv}

\usepackage[utf8]{inputenc} 
\usepackage[T1]{fontenc}    
\usepackage{hyperref}       
\usepackage{url}            
\usepackage{booktabs}       
\usepackage{amsfonts}       
\usepackage{nicefrac}       
\usepackage{microtype}      
\usepackage{lipsum}
\usepackage{graphicx}

\title{Modeling competitive evolution of multiple languages}

\author{
  Zejie Zhou, Boleslaw K. Szymanski, and Jianxi Gao \\
  Department of Computer Science\\
  Rensselaer Polytechnic Institute\\
  Troy, New York 12180, USA \\
  \texttt{gaoj8@rpi.edu} \\
}

\begin{document}
\maketitle

\begin{abstract}
Increasing evidence demonstrates that in many places language coexistence has become ubiquitous and essential for supporting language and cultural diversity and associated with its financial and economic benefits. The competitive evolution among multiple languages determines the evolution outcome, either coexistence,  decline, or extinction. Here, we extend the Abrams-Strogatz model of language competition to multiple languages and then validate it by analyzing the behavioral transitions of language usage over the recent several decades in Singapore and Hong Kong. In each case, we estimate from data the model parameters that measure each language utility for its speakers and the strength of two biases, the majority preference for their language, and the minority aversion to it. The values of these two biases decide which language is the fastest growing in the competition and what would be the stable state of the system. We also study the system convergence time to stable states and discover the existence of tipping points with multiple attractors. Moreover, the critical slowdown of convergence to the stable fractions of language users appears near and peaks at the tipping points, signaling when the system approaches them. Our analysis furthers our understanding of multiple language evolution and the role of tipping points in behavioral transitions. These insights may help to protect languages from extinction and retain the language and cultural diversity.
\end{abstract}

\keywords{Behavioral transition \and Language competition \and Data analysis}

\section{Introduction}
Language is for its speaker an essential component of their culture with great importance also for business and economic activities, especially those involving international knowledge transfer~\cite{welch2008importance}, interdisciplinary research~\cite{bracken2006you}, or international management processes~\cite{welch2005speaking}. The dynamic of the language competition has attracted considerable attention in the past decades, resulting in the development of the mathematical models for competition between two languages~\cite{patriarca2004modeling}, language acquisition, variation across languages~\cite{bates1987competition}, and dynamics of language norm changes~\cite{amato2018dynamics}. Language competitive dynamics fuel the changes in fractions of speakers of various languages~\cite{patriarca2012modeling}, and their collisions~\cite{de2003languages}, blending, and evolution. The final outcome of such evolution can be dominance of one language over the others, the extinction of dominated languages~\cite{abrams2003linguistics}, language coexistence, or unification of close languages into one. This process is affected by both internal and external factors. Internal factors represent inherent characteristics of languages, such as lexical and phonological factors~\cite{costa2003another}. External factors account for social, political and economic influences, such as "The Speak Good English Movement" in Singapore~\cite{suhonen2011speak} and implementation of standardized Mandarin in China~\cite{you2018language}. Both factors influence how people choose their languages and indirectly determine the fraction of speakers of languages, leading to equilibrium with different fractions of speakers for those languages. Languages under the multiple language competitive dynamics may be in one of  three different states: (i) dominate state, i.e., entire population only speak this language; (ii) coexistence state, i.e., there are positive fractions of the population using this language; (iii) extinction state, no one speaks this language.

\noindent
Language dynamics is important for understanding the connections between languages~\cite{tomasello2009constructing} which is related to the language competition and language learning~\cite{bates1987competition}, and second language acquisition~\cite{de2005second} that supports the existence of bilingual interactive activation~\cite{dijkstra1998simulating}. Furthermore, analysis and modeling of language competition can also be extended to social sciences such as population interactions~\cite{pinasco2006coexistence}, formation of collective opinions~\cite{nishi2013collective}, cultural evolution~\cite{michel2011quantitative}, increases language and cultural diversity~\cite{ginsburgh2011many}, and opinion dynamics~\cite{castellano2009statistical}.

\noindent
In the past decade, a variety of models have been developed to understand the competitive dynamics of languages. Most of the attention has been paid to the coexistence of two languages~\cite{pinasco2006coexistence, patriarca2012modeling, stauffer2007microscopic} and the corresponding bilingualism~\cite{sole2010diversity}. One research in this area is finding factors that affect the evolution of competing languages, such as social interaction networks~\cite{patriarca2012modeling}, or microscopic competition between two languages~\cite{stauffer2007microscopic}. Another research area is modeling abstract competition factors by using real data to estimate the model parameters for transforming language from old to new forms~\cite{lieberman2007quantifying} or estimating all parameters in language competition~\cite{zhang2013principles}. Moreover, some researches focus on combining the area of language dynamics with the area of statistical physics~\cite{castellano2009statistical,loreto2011statistical} or applying to them statistical laws to describe word uses~\cite{petersen2012statistical}. These previous works develop approaches to modeling, analyzing, and even quantifying the competition and evolution of languages, enabling us to theoretically construct and simulate dynamics of two languages. In~\cite{fujie2013model}, the authors extend the Abrams-Strogatz model to the competition among multiple languages, but their model becomes very complicated with large number of languages. Still, this multiple language competition model has not been validated on real-word data. Thus, we ask the critical questions: What are the critical parameters that determine the existence and values of tipping points of language coexistence and extinction in the community? And is there a system metric that can indicate the approaching extinction of a language?

\noindent
Here, we answer the questions raised above by using real-world language evolution data from Singapore~\cite{cavallaro2014language} and Hong Kong~\cite{languages, LanguageHongKong, bacon1998charting} to find optimal parameters and validate the extended Abrams-Strogatz model~\cite{fujie2013model}. The model parameters found in this process drive the utilities of competing languages, the strength of majority preference for the most popular language and minority aversion to this language and measure this utility's impact on language evolution.  Currently, all the modeled languages coexist. We investigate the behavioral transitions of the languages under the perturbations of each parameter. We find that the language with the highest language utility tends to grow faster and eventually gain the largest fraction of speakers. Moreover, when the majority preference is small than a certain critical value, the popular languages may lose its leading position during the evolution process. Finally, when the minority aversion is not sufficiently strong, the languages with small (including the language with the smallest) initial fractions of speakers may gain the largest fraction of speakers. From the above analysis, we obtain the complete phase diagram for each community, showing the evolution of each language in each dataset and the relation between transitions and parameters. Secondly, we analyze the relation between convergence time and state of competing languages, and show that the competition arises to the highest level when the language dominance switches from one language to another. Finally, we illustrate individual and combined effects of two language biases, the majority preference and the minority aversion, by simulating how languages with the largest initial fractions of speakers are affected.

\section{RESULT}
    \subsection{DATA AND MODELING}

To model the real world language competitions, we use dataset of languages used in Singapore (the whole country), languages used in the Chinese community of Singapore, languages used in Indian Community of Singapore, and languages used in Hong Kong. We consider speakers of one language in our dataset as people who consider this language as their primary language.

\noindent  
    \textbf{Singapore.} In the 1950s, dialects such as Hokkien were the most widely spoken language in Singapore. In the 1957 census, about 1.8\% people mainly spoke English, about 0.1\% people mainly spoke Chinese Mandarin. However, after the implementation of a series of policies from the 1950s to present, the proportions of speakers of different languages in Singapore considerably changed. Until 2010, English and Chinese Mandarin became the most spoken languages with speakers proportion of 32.3\%, 35.6\%, respectively in Singapore's entire country~\cite{cavallaro2014language}. 

\noindent    
    \textbf{Hong Kong.} We use language data collected between 1949 to 2016 in Hong Kong~\cite{LanguageHongKong,bacon1998charting,languages}. The number of people who mainly speak English increased in these 67 years, surpassing the number of people who mainly speak Hakka, Hoklo, or Sze Yap. We do not include Cantonese speakers in our dataset since the number of people who use Cantonese as their common language is much larger than the speaker population in our dataset. We normalize all languages in our dataset before fitting then into our model.

\noindent	
	We employ the extended Abrams-Strogatz model~\cite{fujie2013model} to test the competition among multiple languages,
	\begin{equation} \label{eq:1}
        \frac{dx_{i}}{dt} = \sum^{n}_{j=1, j\neq i}x_{j}P_{ji} - x_{i}\sum^{n}_{j=1, j\neq i}P_{ij}
    \end{equation}

\noindent	
where $x_{i}$ is the fraction of the population speaking language $i$, and $P_{ij}$ represents the transition rate from language $j$ to language $i$.
    \begin{equation} \label{eq:2}
        P_{ij} = s_{i}x_{i}^\beta(1-x_{j})^{\alpha - \beta}
    \end{equation}

\noindent    
where $\beta (\geq 0)$ and $\alpha - \beta (\geq 0)$ represent the strength of the majority preference and the minority aversion, respectively. $s_{i}>0 $ is the utility of language $i$, and   $\sum_{i=1}^{n} s_{i}=1$.

\noindent
We utilize numerical simulation to compute the parameters in our model. The following equation calculates the difference:
     \begin{equation} \label{eq:3}
        D = \sum_{j=1}^{m}\sqrt{\sum_{i = 1}^{n}(x_{i,j} - x_{i,j}')^2} 
    \end{equation}

\noindent
    where $i$ is the time step (we use year as a time step unit) varying from 1 to n which is the number of total time steps, $j$ is the index of languages varying from 1 to m, which is the number of languages. $x_{i,j}$ is the rational value of fraction of language j users at time step i, $x_{i,j}'$ is the theoretical value of a fraction of language j users in time step i. 

\noindent    
    To fit this model to our language data, we set a range for parameter majority preference $\beta$ and minority aversion $\alpha - \beta$, and then, iterating over this range of majority preference and minority aversion, we evaluate possible values for language utilities $s_{1}, s_{2}...s_{m}$. Given $\beta$ and $\alpha - \beta$ to minimize the difference $D$ between real language fractions and theoretical language fractions we calculate from Eq.~\ref{eq:1}. Then we repeat narrowing range of parameters and doing grid--search for language utilities, $\beta$, and $\alpha - \beta$ to keep increasing the precision of our theoretical model. We end the repetition until our theoretical model is close enough to the real world language data. Note that since parameter $s_{i}$, which represents state utility in extended Abrams-Strogatz model, acts as a useful or advantageous factor or feature of state $i$, in our language competition model, we change the definition of $s_{i}$ to the utility of language $i$.
    

    \begin{figure}
    \centering
    \includegraphics[width=\textwidth]{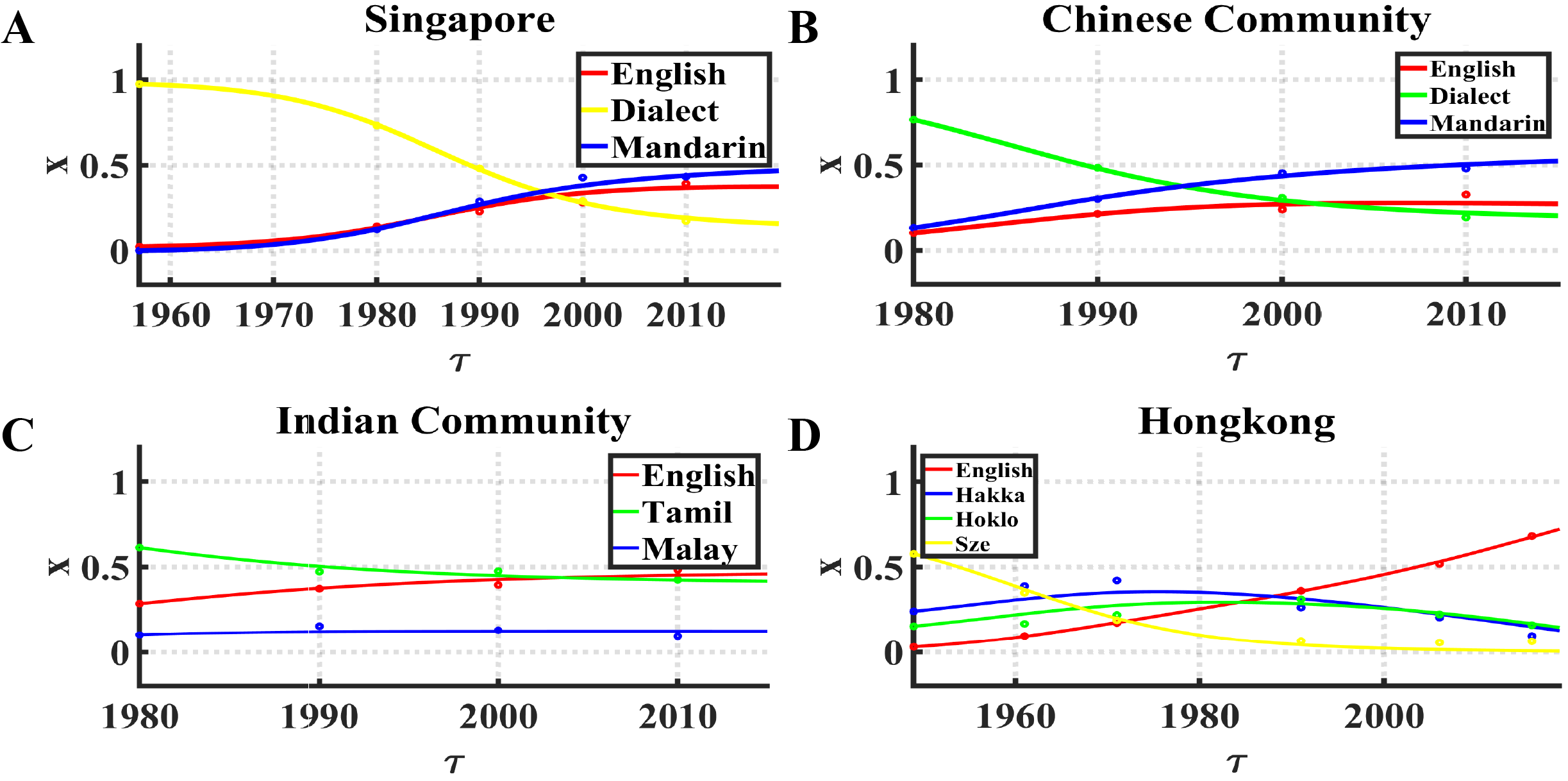}
    \caption{{\bf The language evolution over time.} Real data are shown by dots, theoretical data are shown by solid lines, real and theoretical data are all normalized. X represents the fraction of language, $\tau$ represents the time point. (A) Language data of the whole country in Singapore between 1957 and 2010. Parameters fit to dynamic model ($\alpha=1.009$, $\beta = 0.726$, English: $s_{i}=0.35$, Dialect: $s_{j}=0.29$, Mandarin: $s_{k} = 0.36$), error is $D=0.1388$. (B) Language data of Chinese community in Singapore between 1957 and 2010. Parameters fit to dynamic model($\alpha=0.99$, $\beta = 0.63$, English: $s_{i}=0.33$, Dialect: $s_{j}=0.3$, Mandarin: $s_{k} = 0.37$), error is $D=0.1199$. (C) Language data of Indian community in Singapore between 1957 and 2010. Parameters fit to dynamic model ($\alpha=1.03$, $\beta = 0.21$, English: $s_{i}=0.4$, Tamil: $s_{j}=0.39$, Malay: $s_{k} = 0.21$), error is $D=0.1323$. (D) Language data of Hong Kong between 1949 and 2016. Parameters fit to dynamic model ($\alpha=1.082$, $\beta = 0.987$, English: $s_{i}=0.307$, Hakka: $s_{j}=0.252$, Hoklo: $s_{k} = 0.263$, Sze Yap: $s_{l} = 0.178$), error is $D=0.2663$ }
    \label{modelFit}
    \end{figure}
    \noindent
    By fitting the language data into extended Abrams-Strogatz model, we compare the real evolution processes in different districts and their corresponding simulation results in Fig.~\ref{modelFit}. It is clear that the model successfully represents the behavior of the data, correctly showing the evolution of each language. In Singapore, the whole country dataset, Dialect starts with about 0.975 fractions of speakers and is surpassed by Mandarin in 1996, by English in 1997. As for Chinese community of Singapore dataset, the trend of speakers in a different language is similar to that of speakers in Singapore the whole country dataset where English and Mandarin gradually replace dialects: it starts with about 0.766 fractions of speakers and is surpassed by Mandarin in 1994, by English in 2001. In the Indian community of Singapore dataset, Tamil started with the most substantial fraction (about 0.613) of speakers but continually lost its speakers and eventually was exceeded by English in 2003. The commonality of these three datasets is the increase of English speakers, which might be caused by the increasing number of English medium schools~\cite{cavallaro2014language} in this period. The increasing usage of Mandarin in Singapore the whole country dataset and Chinese community dataset might benefit from "Speak Mandarin Campaign" implemented in 1979 in Singapore~\cite{leong2014study}. In Hong Kong dataset, Sze Yap, which is a Chinese vernacular in Hong Kong, owns the most substantial fraction of speakers(about 0.578) at the beginning, but is gradually replaced by the remaining three languages, and it almost goes to extinction in 1999.

\subsection{STATE DEFINITION}
We will refer to the language with the largest number of speakers as the most popular, but if this language drives the competing languages to extinction, we will refer to is as dominant.

\begin{figure}
    \centering     
    \includegraphics[width = \textwidth]{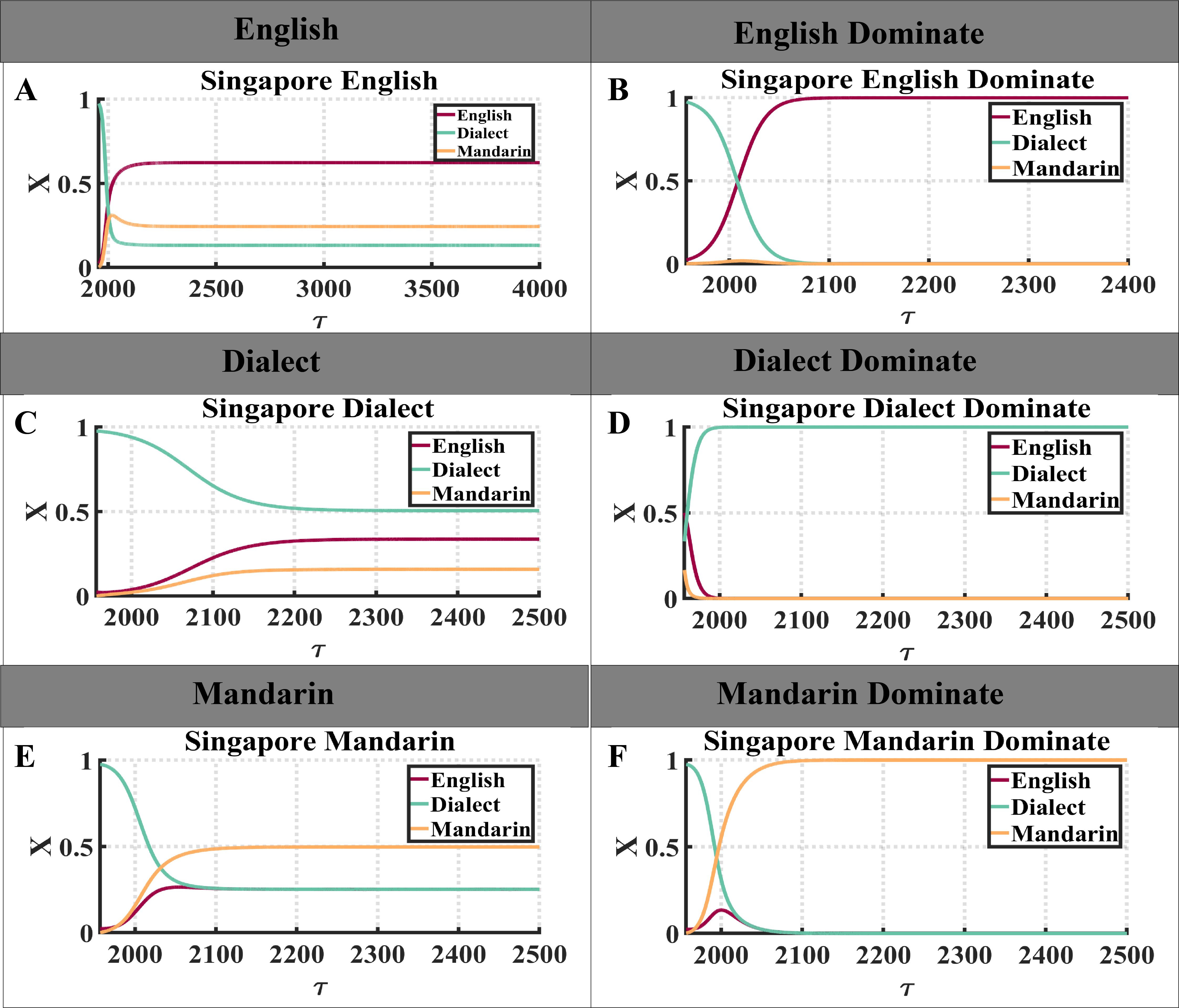}
    \caption{We take Singapore dataset to illustrate "coexistence state" and "dominance state" of a language. "X" represents the fractions of languages and $\tau$ represents time steps. (A) After $\tau = 2500$, competing languages stay in coexistence state with English acquiring the largest fraction of speakers. (B) English is dominant since from $\tau = 2100$, the fractions of all other languages decrease to an extremely low level. (C) Dialect gets the largest fraction of speakers without leading to the extinction of Dialect or Mandarin. (D) Dialect's fraction of speakers increase to an extremely high level, leading to the extinction of other languages. (E) Mandarin's fraction of speakers grows and surpass English and Dialect. Eventually, all languages stay in coexistence state and Mandarin owns the largest fraction of speakers. (F) As Mandarin increases its number of speakers significantly, other competing languages go extinct.}
    \label{definition}
    \end{figure}

\noindent    
    Here we need to define each state in the evolution of language competition. In our simulations, two distinct states are defined as: "coexistence state" and "dominance state". "Coexistence state" arises if at least two languages survive but in this state, an extinction of some languages is still possible. In contrast, in "dominance state", the survival of one language leads to the extinction of all others. The "coexistence state" and "dominance state" are illustrated in Fig.~\ref{definition}, where the left column shows example of "coexistence state" with its title showing the language with largest fraction of speakers, while the right column displays examples of "dominance state". Note that the "coexistence state" and "dominance state" here all refer to the state after fractions of speakers for all languages converge to their steady states. 

\noindent    
   As shown in Fig.~\ref{definition}A, during language competition, Dialect's fraction of speakers drops dramatically, while English's fraction of speakers increases. Eventually, none of the competing languages disappear, with English being the most popular language. In contrast, English is shown to be the only survivor after the language competition in Fig.~\ref{definition}B, because English gains a substantial fraction of speakers while other languages lose almost all theirs. In Fig.~\ref{definition}C, Dialect loses a large number of speakers in the competition, causing its competitors to gain more speakers, but it still holds the most speakers and coexists with other competing languages. Dialect surpasses English and becomes dominant in Fig.~\ref{definition}D, where fractions of speakers of languages other than Dialect all reduce to a shallow level. Similarly in Fig.~\ref{definition}E, Mandarin owns the largest fraction of speakers and stays in "coexistence state," while in Fig.~\ref{definition}F, Mandarin leads to the extinction of all other languages. As we took the Singapore dataset as an example to define different states, we are now able to describe future simulations in a more straight forward way. 

\subsection{STATE UTILITY $s_{i}$}
    In the extended Abrams-Strogatz model, $s_{i}$ represents the utility of language $i$. Accordingly, in our language dataset, each language $i$ has its own $s_{i}$ representing its utility. We analyze the relation between $s_{i}$ and the competition between language $i$ and other languages when $s$ is in the range $[0, 0.6]$. Since the total of all language utilities is by definition 1, in each simulation, with the increasing of the utility of one language, increasing utility of one language decreases utilities of other languages proportionally to their current utility.
    \begin{figure}
    \centering     
    \includegraphics[width = \textwidth]{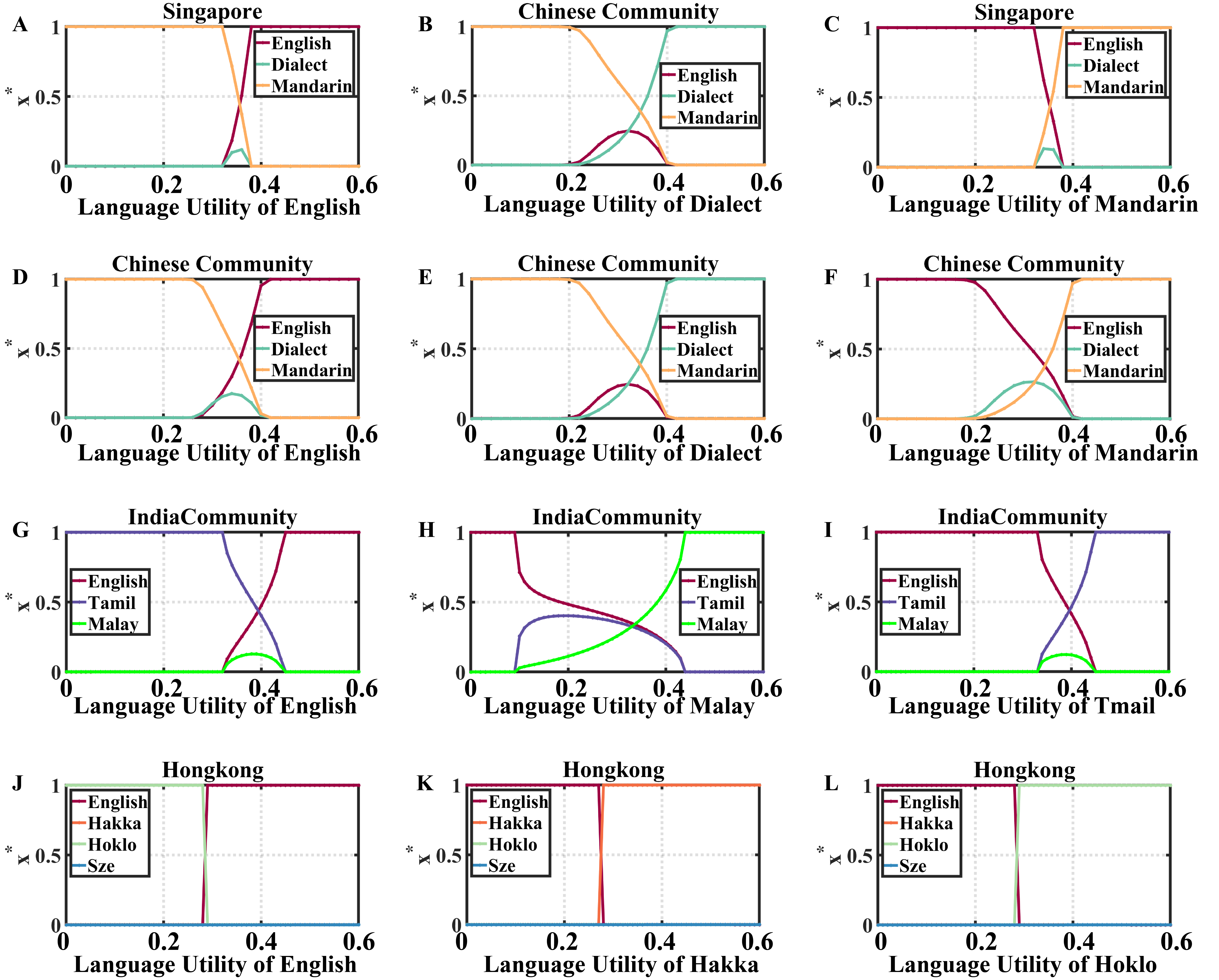}
    \caption{Fraction of language speakers in relation to this language utility, where $X^{*}$ represents the fraction of a language users at the steady state of competition, and $s$ stands for language utility. (A) Fraction of language speakers in relation to the English utility in the Singapore whole country dataset. (B) Fraction of language speakers in relation to Dialect utility in the same dataset. (C) Fraction of language speakers in relation to Mandarin utility in the same dataset. (D) Fraction of language speakers in relation to English utility in Chinese community in Singapore. (E) Fraction of language speakers in relation to Dialect utility in Chinese community in Singapore. (F) Fraction of language speakers in relation to Mandarin utility in Chinese community in Singapore. (G) Fraction of language speakers in relation to English utility in Indian community in Singapore. (H) Fraction of language speakers in relation to  Tamil utility in Indian community. (I) Fraction of language speakers in relation to Malay utility in Indian community. (J) Fraction of language speakers in relation to English utility in Hong Kong. (K) Fraction of language speakers in relation to  Hakka utility in Hong Kong. (L) Fraction of language speakers in relation to Hoklo utility in Hong Kong. }
    \label{controlS}
    \end{figure}
    \noindent
     In the Singapore whole country dataset (Fig.~\ref{controlS}ABC), we set majority preference $\beta = 0.726$ and minority aversion $\alpha - \beta = 0.283$. For Fig.~\ref{controlS}A, when $s_{i}\in [0, 0.32]$, languages are in "dominance state" with Mandarin being dominant. As $s_{i}$ further increases, languages come into "coexistence state" for a short while, and the fraction of English speakers surpasses the fraction of Mandarin speakers for $s_{i}>0.34$. When $s_{i}>0.36$, English becomes dominant and drives others to extinction. In Fig.~\ref{controlS}B, language competition starts in "dominance state" as well (language utility of Dialect $s_{j}\in [0, 0.24]$), with Mandarin being dominant. Then, the system reaches "coexistence state" in which all competing languages' fractions change dramatically. When the utility of Dialect exceeds 0.36, it becomes dominant. Similarly, in Fig.~\ref{controlS}C, in the beginning, when  Mandarin utility $s_{k}\in[0, 0.32]$, English is dominant. Then, for $s_{k}\in [0.34, 0.36]$, the system is in "coexistence state" and the fraction of English speakers drops significantly. Eventually, Mandarin becomes dominant when its utility exceeds 0.36. 

\noindent   
    In Chinese community dataset (Fig.~\ref{controlS}DEF), we set majority preference $\beta = 0.63$ and minority aversion $\alpha - \beta = 0.36$. In Fig.~\ref{controlS}D, Mandarin is dominant when $s_{i} \in [0, 0.3]$. Then the system reaches "coexistence state" when $s_{i}$ increases. With further increase of $s_{i}$, Dialect becomes dominant. In Fig.~\ref{controlS}E, when language utility of Dialect $s_{j}\in [0, 0.2]$, Mandarin is dominant. When $s_{j}\in [0.22, 0.4]$, the system is in "coexistence state". When $s_{j}>0.4$, Dialect becomes dominant. In Fig.~\ref{controlS}F, when language utility of Mandarin $s_{k}\in [0, 0.18]$, English is dominant. When $s_{k}\in [0.2, 0.4]$, the system is in "coexistence state". When $s_{k}>0.4$, Mandarin becomes dominant. 
     
     \noindent
     In Indian community dataset (Fig.~\ref{controlS}GHI), we set majority preference $\beta = 0.21$ and minority aversion $\alpha - \beta = 0.82$. In Fig.~\ref{controlS}G, when language utility of English $s_{i}\in [0, 0.32]$, Tamil, which is the language with the largest utility,  is dominant. Then the system reaches "coexistence state" as the utility of English further increases. When $s_{i}$ exceeds 0.44, English starts to be dominant. In Fig.~\ref{controlS}H, when language utility of Tamil $s_{j}\in [0, 0.33]$, English is dominant. When $s_{j}\in [0.34, 0.45]$, the system is in "coexistence state". When $s_{j}>0.45$, Tamil becomes dominant. In Fig.~\ref{controlS}I,  English is dominant for the initial range of $s_{k}$, then its fraction of speakers decrease significantly, causing other languages to grow as the system enters "coexistence state." When $s_{k}>0.43$, Malay is dominant. 
     
     \noindent
     In Hong Kong dataset (Fig.~\ref{controlS}JKL), we set majority preference $\beta = 0.987$ and minority aversion $\alpha - \beta = 0.095$. In Fig.~\ref{controlS}J, when language attraction of English $s_{i}\in [0, 0.28]$, Hoklo is dominant. Then Hoklo and English's fractions of speakers change dramatically, leading to the dominance of English. In Fig.~\ref{controlS}K, when language attraction of Hakka $s_{j}\in [0, 0.27]$, English is dominant. Then, as language utility of Hakka exceeds 0.27, Hakka becomes dominant. Similarly, in Fig.~\ref{controlS}L, in the beginning, English is dominant, but for $s_{k}>0.28$, Hoklo becomes dominant.
     
     \noindent
     From these simulations, we find that when the language utility $s_{i}$ of language $i$ is relatively small, one of the other languages, which is usually the language with the highest language utility, tends to be dominant. As $s_{i}$ increases, languages might come into "coexistence state," which acts as a transition period for language $i$ to become dominant. Moreover, when $s_{i}$ is large enough, language $i$ becomes dominant.

\subsection{CONVERGENCE TIME}
\begin{figure}[tp]
    \centering     
    \vspace*{-1cm}
    \includegraphics[width=\textwidth]{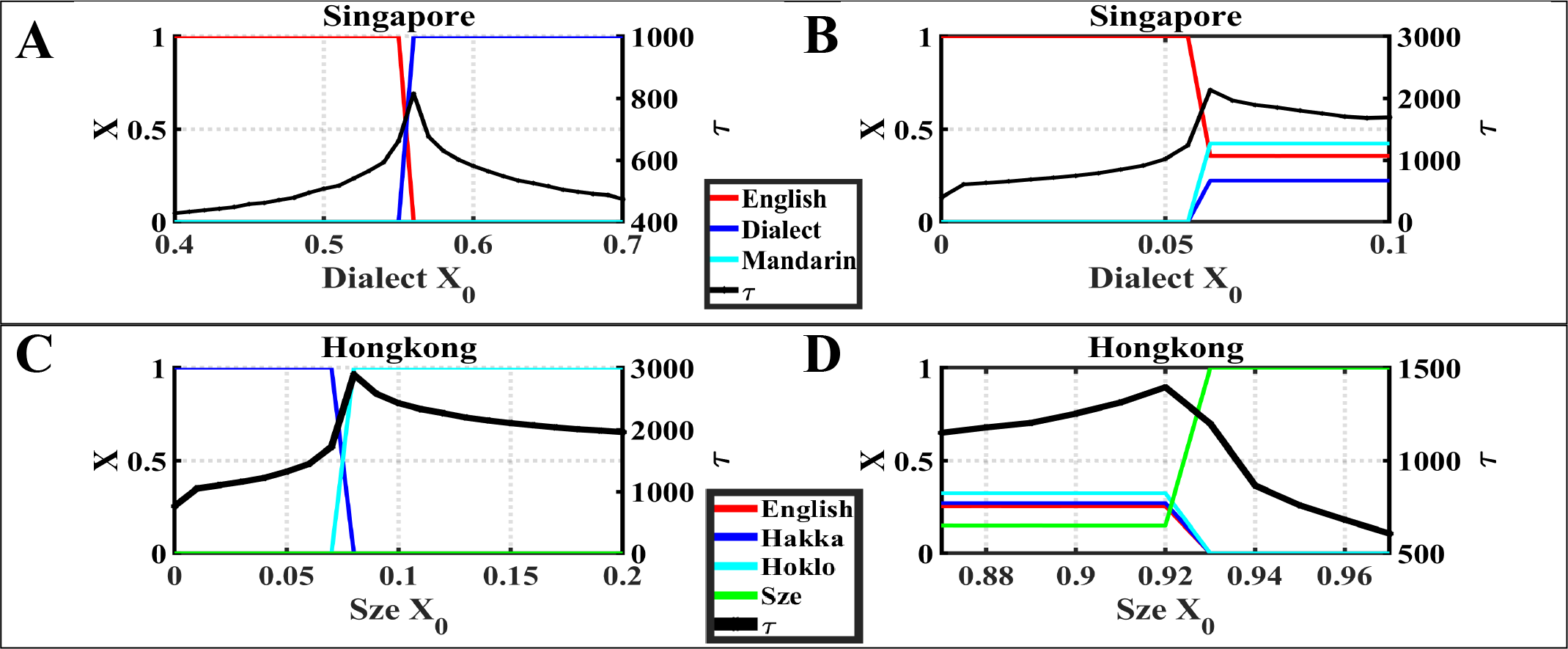}
    \caption{We gradually increase the initial fraction of one language to observe the relation between times used for all languages to reach the steady state of their fractions of speakers and the distance to the tipping point for the initial fraction. (A) When the language dominance switches from one language to another, time to achieve the steady state of each language speakers' fractions reaches a peak. (B) Similar to subfigure (A), in Hong Kong dataset, when the language dominance switches from one language to another, time to achieve the mentioned above steady state reaches a peak. (C) When the system transitions from 'coexistence state' to 'dominance state', time to achieve the mentioned above steady state again reaches a peak. (D) Before the system transitions from 'dominance state' to 'coexistence state', time to achieved the mentioned above steady state  reaches a peak.}
    \label{convergence}
    \end{figure}

  \noindent
    It is notably hard to predict the critical transition from one state to another because the state of the system may show little change before the tipping point~\cite{scheffer2009early}. Critical slowdown~\cite{scheffer2001catastrophic} defined in statistical physics is an indicator for early warning signals with applications to many fields, ranging from the economy~\cite{diks2015critical} to ecology~\cite{lenton2012early}. Here we employ the convergence time as the early warning signals for the behavioral transition in the language competition. Fig.~\ref{convergence} shows the convergence time of different datasets under different language competitions, where the $x$-axis represents the initial fraction of one language, the left $y$-axis represents the equilibrium fraction of each language, and the right $y$-axis represents the convergence time $\tau$. In Fig.~\ref{convergence}A, $\tau$ reaches a peak when the initial fraction of Dialect increases to 0.56 and Dialect replaces English as dominant language. Similarly, at $\tau$ in Fig.~\ref{convergence}B, when the system transitions from "dominance state" to "coexistence state", $\tau$ reaches its peak. In Hong Kong datasets, we find similar outcome. The convergence pattern observed in Fig.~\ref{convergence}C is similar to the one seen in Fig.~\ref{convergence}A since the peak of $\tau$ happens when the dominance switches from from one language to another. As for Fig.~\ref{convergence}D, the peak of $\tau$ happens when the system transitions from "coexistence state" to "dominance state," which is exactly opposite to the transition in Fig.~\ref{convergence}B, yet they show similar patterns of convergence time. 
  
   \noindent  
 From these simulations, the peak of convergence time happens when the state transition happens of either switching the dominant language or moving from "coexistence state" to "dominance state," or vice versa. Such transitions can be caused by the comparable competing strength of different languages. The convergence time enables us to identify the "tipping point" of the system parameter values giving the control system enough time to prevent unwanted transition.

\subsection{SENSITIVITY TO MAJORITY PREFERENCE AND MINORITY AVERSION}
    \noindent
   Here, we focus on how majority preference and minority aversion can affect language competitions. For each dataset, we set all parameters and languages' initial fractions of speakers to the values used in modeling and data section except for majority preference $\beta$ and minority aversion $\alpha - \beta $. 

\begin{figure}
    \centering     
    \vspace{-1cm}
    \includegraphics[width = \textwidth]{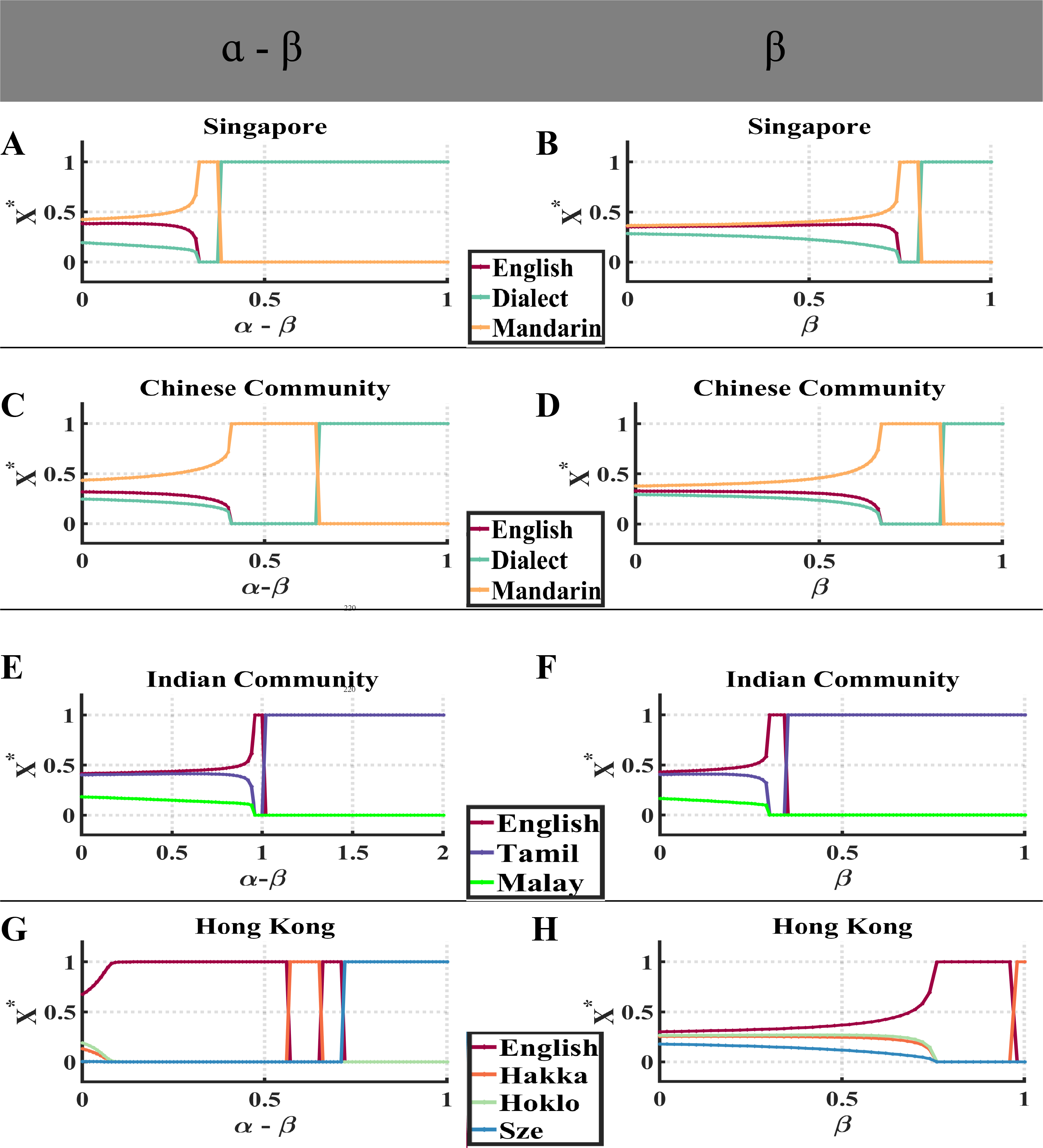}
    \caption{The change of equilibrium point when $\alpha - \beta$ or $\beta$ change in each language dataset. (A) The relation between $\alpha - \beta$ and fraction of each language in the Singapore whole country dataset. (B) The relation between $\beta$ and fraction of each language in the Singapore whole country data set. (C) The relation between $\alpha - \beta$ and fraction of each language in Chinese community in Singapore. (D) The relation between $\beta$ and fraction of each language in Chinese community in Singapore. (E) The relation between $\alpha - \beta$ and fraction of each language in Indian community in Singapore. (F) The relation between $\beta$ and fraction of each language in Indian community in Singapore. (G) The relation between $\alpha - \beta$ and fraction of each language in Hong Kong. The relation between $\beta$ and the fractions of each language speakers in Hong Kong.}
    \label{single}
    \end{figure}

In Fig.~\ref{single}A, we have $\beta = 0.726$ and $\alpha - \beta$ ranging from 0 to 1 with step 0.01. In the range of $\alpha - \beta\in[0, 0.31]$, Mandarin has the largest fraction of speakers, and all languages are in "coexistence state". In the range of $\alpha - \beta\in[0.32, 0.37]$, Mandarin is dominant, causing the extinction of Dialect and English. When the minority aversion start to exceed 0.38, then Dialect, which is the language with the largest initial fraction in this dataset, starts to be dominant. In Fig.~\ref{single}C, in the range of $\alpha - \beta\in[0, 0.4]$, the system stays in "coexistence state" and English is dominant. When $\alpha - \beta$ further increases, English still is dominant, but as the minority aversion becomes large enough, language with the largest initial fraction (Dialect) of speakers starts to be dominant. In Fig.~\ref{single}E, the system starts in "coexistence state" and English, which is the language with the largest language utility ($s=0.4$), is dominant in the range of $\alpha - \beta\in[0, 0.96]$. Then, when $\alpha - \beta\in[0.98, 1.02]$, English is again dominant. Tamil, the language with the largest initial fraction of speakers, is dominant when its minority aversion is larger than 1.02. In Fig.~\ref{single}G, when $\alpha - \beta\in[0, 0.07]$, English, which is the language with the largest language utility ($s = 0.297$), is most popular, but the system is in "coexistence state". When $\alpha - \beta\in[0.08, 0.56]$, English is again dominant. When $\alpha - \beta\in[0.57, 0.71]$, Hakka and English compete with each other for dominance, with Hakka is dominant when $\alpha - \beta\in[0.57, 0.65]$ and English is dominant when $\alpha - \beta\in[0.66, 0.71]$. This may be caused by the two languages' similar level of competitiveness determined by both initial fraction and utility. However, when minority aversion is high enough (larger than 0.71), Sze Yap, which is the language with the highest initial fraction, becomes dominant. 

From these simulations, we conclude that low value of minority aversion favors the growth of languages with small initial fraction, and usually the language with the largest language utility $s$ among them, can gain the largest fraction of speakers. In contrast, high minority aversion favors the growth of languages with the largest initial fraction of speakers.
    
  \noindent   
    in Fig.~\ref{single}B we set the minority aversion $\alpha - \beta = 0.28$, and vary the majority preference $\beta$ from $[0, 1]$.  Similar to language patterns in Fig.~\ref{single}A, languages start in "coexistence state", with Mandarin owning the largest fraction of speakers when $\beta\in [0, 0.74]$. When the majority preference increases to ($\beta\in [0.75, 0.8]$), Mandarin becomes dominant, even though it had the smallest initial fraction of speakers. As the majority preference further increases and until it reaches 0.82, Dialect, the language with the largest initial fraction, starts to be dominant. 

In Fig.~\ref{single}D, when $\alpha - \beta = 0.36$, the system stays in "coexistence state", with Mandarin being dominant in the range of $\beta\in[0, 0.66]$. Then as $\beta$ increases to ($\beta\in [0.67, 0.83]$), Mandarin is dominant for a short range of $\beta$ because for $\beta>0.83$, Dialect, the language with the smallest initial fraction, is dominant. In Fig.~\ref{single}F, at the beginning with ($\beta\in[0, 0.3]$), English, which is the language with the largest language utility, has the largest fraction of speakers and all languages are in "coexistence state." For $\beta\in [0.31, 0.34]$, English is dominant, but for larger $\beta$, Tamil becomes dominant. Similarly, as shown in Fig.~\ref{single}H, the system transitions from "coexistence state" to "dominance state", and the language with the largest initial fraction becomes dominant when majority preference is high enough; in this case, for $\beta\in[0.76, 0.96]$ English is dominant, while for $\beta\in[0.98, 1.16]$,  Hakka takes over this role. Finally, when $\beta>1.16$, Sze Yap, the language with the largest initial fraction, is dominant. 

From these four simulations, we find that when majority preference is small, languages with small initial fraction (usually the language with the largest language utility) might own the largest fraction of speakers, and becomes dominant. However, when majority preference is high enough, the language with the largest fraction of speakers usually becomes dominant, driving other language to  extinction.
    
    \begin{figure}[]
    \centering     
    \vspace{-1cm}
    \includegraphics[width = \textwidth]{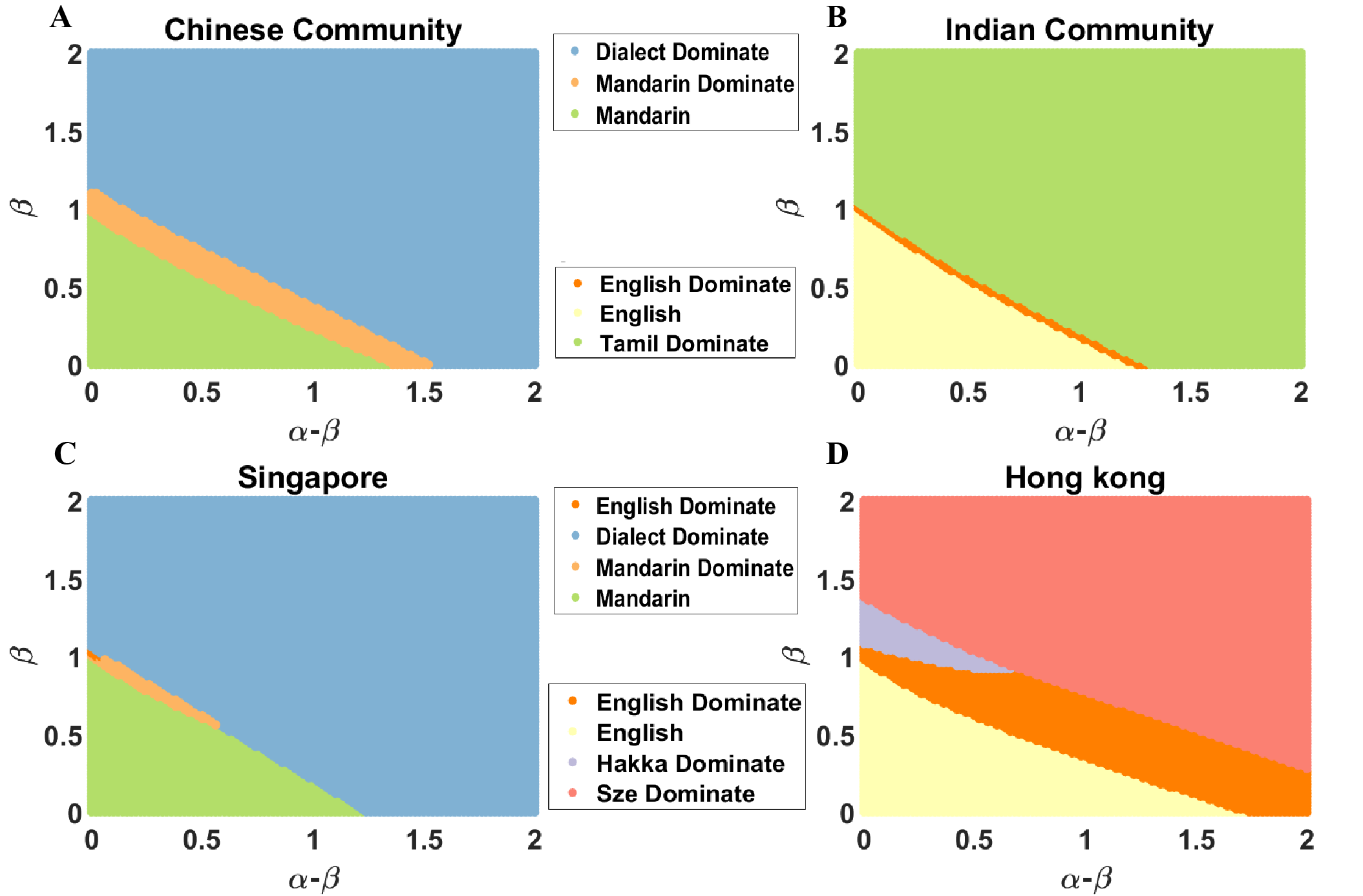}
    \caption{States of the language with the largest fraction of speakers as a function of increasing the majority preference and the minority aversion. "Dominate" denotes one language domination over others, otherwise, all languages shown in the figure coexist. (A) The language with the largest fraction of speakers in relation to $\alpha - \beta$ and $\beta$ in Chinese community in Singapore. (B) The language with the largest fraction of speakers in relation to $\alpha - \beta$ and $\beta$ in Indian Community in Singapore. (C) The language with the largest fraction of speakers in relation to $\alpha - \beta$ and $\beta$ in the Singapore whole country data set. (D) The language with the largest fraction speakers in relation to $\alpha - \beta$ and $\beta$ in Hong Kong.}
    \label{double}
    \end{figure}
\subsection{PHASE DIAGRAM}
 
  \noindent    
   As for the combined effect of majority preference and minority aversion, we still use the language initial fraction of speakers and parameters from the data and modeling section. Here, we only consider the language with the largest fraction of speakers. 
In Fig.~\ref{double}A, when majority preference $\beta \in [0, 0.98]$, three different kinds of patterns appear. In the first pattern, all three languages (English, Dialect, and Mandarin) coexist and Mandarin is the most popular. The next pattern has Mandarin dominant. The third pattern has Dialect dominant. In this case, when minority aversion $\alpha - \beta$ is large enough, Dialect, which is the language with the most speakers, is dominate. As $\beta$ increases from 0 to 0.98, Dialect becomes more and more likely to become dominant because the range of $\alpha - \beta$ for Dialect to be in this role  becomes larger and larger. When $\beta \in [0.98, 1.1]$, two different patterns arise. In the first, Mandarin is dominant while in the second, it is Dialect which is dominant. When $\beta$ exceeds 1.1, Dialect is dominant. Fig.~\ref{double}B shows that for the majority preference $\beta\in [0, 1]$, again three different patterns arise. In the first one, three languages (English, Tamil, Malay) coexist and Tamil is the most popular. In the second pattern, English is dominant, while in the third it is Tamil that is dominant. Similar to Fig.~\ref{double}(A), the language with the largest utility (Mandarin in Fig.~\ref{double}A and English in Fig.~\ref{double}B) tends to be the most popular when the majority preference and minority aversion are small. With $\beta \in [1, 1.02]$, two different patterns are present. In the first, again three languages coexist, and English is the most popular. The second has English dominant. The third pattern arises when $\beta>1.02$, Tamil, which is the language with the largest initial fraction of speakers, is dominant. 
 
\noindent     
Fig.~\ref{double}C shows that when minority aversion $\alpha - \beta \in [0, 0.04]$, four different patterns arise. In the first, again three languages (English, Dialect, Mandarin) coexist and Mandarin has the most speakers. The second has Mandarin is dominant, while in the third pattern it is English that is dominant. Finally in the fourth case Dialect is dominant. In this case, even English can  be dominant over a short range of parameters, because of its comparably high initial fraction and language utility; it has the second largest initial fraction and the second largest language utility. With $\alpha - \beta \in [0.06, 0.56]$, the first three patterns from the case of lowest $\beta$ reappear. Again, Mandarin is the language with the largest utility and it is the most when minority aversion and majority preference are small. When $\alpha - \beta$ is greater than 0.56, Dialect is dominant. In Fig.~\ref{double}D, with minority aversion $\alpha - \beta \in [0, 0.66]$, four different patterns arise. In the first, four languages (English, Hakka, Hoklo, Sze Yap) coexist, and English is the most popular. In the second, English is  dominant while in the third it is Hakka that is dominant. Finally, in the fourth, Sze Yap is dominant. When $\alpha - \beta \in [0.68, 2]$, the previous patterns, the first, the second, and the fourth reappear. In this dataset, English has the largest language utility and owns the largest fraction of speakers when the majority preference and minority aversion are small. 

 \noindent     
    When the majority preference and the minority aversion are relatively small, competing languages tend to coexist with each other and language with the largest utility tends to own the most speakers. Hence, when the majority preference and the minority aversion are relatively small, they affect language competition weakly. It is the language utility that plays an essential role in this competition. As the majority preference and the minority aversion increase, some languages become dominant. 
When the majority preference and the minority aversion further increase, the language with the largest initial fraction is dominant, indicating that when the majority preference and the minority aversion are large enough, they will favor the growth of the language with the largest initial fraction of speakers and make it dominant.
  
\section{DISCUSSION}
	Here we provide a model and its validation using four real-world language competitions involving several languages that extends the Abrams-Strogatz model in the important direction. The model fits well with the real data, as shown in the first section, enabling us to further analyze factors affecting the language competition in detail. Our contributions can be summarized as follows. 

 \noindent     
    1. We show that language utility affects the competitive evolution of communities using several languages. When the utility of a language is low enough, it might go extinct, but when this utility is high enough, it can be dominant and drive other languages to extinction. However, it is also possible that with the value of this utility in mid-range the system can be in a "coexistence state". 

 \noindent     
    2. The relation between convergence time and state transition of languages shows that convergence time to steady state fractions of language users reaches a peak at the state transition tipping points. Such critical slowdown can be caused by similar competitiveness of the different competing languages. At the tipping points either one dominant language is replaced by another, or the system transitions from a "dominance state" to "coexistence state" or vice versa.
 
  \noindent    
    3. We demonstrate the influence that the majority preference and the minority aversion can separately exert on competing languages. When majority preference is small, a  language with small initial fraction of speakers (including the language with the smallest initial fraction of speakers) can have the most speakers after the language competition, and even lead to the extinction of other languages. When the majority preference is large enough, the language with the largest initial fraction of speakers will win the language competition, usually driving  all other languages to extinction. The simulations with varying the minority aversion yield similar results as described above.  

 \noindent     
    4. We discuss also the influence that the majority preference and the minority aversion can together exert on the evolution of competing languages. When both of these biases are relatively low, a language with the small initial fraction of speakers can gain the most speakers in the steady state, and become dominant. When both biases are high enough, the language with the most speakers initially is most likely to be dominant. Moreover, there are variants of these conditions yielding results other than the two reported above, as shown in Fig.~\ref{double}C where English is dominant as shown in Fig.~\ref{double}D where Hakka is dominant. 
 
  \noindent    
Our simulations illustrate the results of the language competition under various conditions, providing examples of the impact of the language utility, the majority preference, and the minority aversion on the competition outcomes. Yet, analytical formulas defining quantitatively the competitive evolution of languages as a function of time and different parameters of the model are not known yet. Moreover, our simulations are completed without considering geographical~\cite{patriarca2009influence}, physiological~\cite{kuhl2011early}, and other factors. Hence, constructing a conclusive formulation of the language competition in the real world requires futures future research.

\section{DATA ACCESSIBILITY}
All data needed to evaluate the conclusions in the paper are present in the paper itself or are available upon request from the authors. 
\section{AUTHOR'S CONTRIBUTIONS}
Z.Z., B.K.S and J.G. conceptualized the study; Z.Z. simulated extensively; Z.Z., B.K.S and J.G. wrote the paper.
\section{COMPETING INTERESTS}
We declare we have no competing interests.
\section{ACKNOWLEDGEMENTS}
This study was supported in part by the Office of Naval Research (ONR) Grant No. N00014-15-1-2640, and by the Army Research Office (ARO) Grant No. W911NF-16-1-0524.

\bibliographystyle{unsrt}
\bibliography{citation} 

\usepackage{CJK}

@article{lenton2012early,
  title={Early warning of climate tipping points from critical slowing down: comparing methods to improve robustness},
  author={Lenton, TM and Livina, VN and Dakos, V and Van Nes, EH and Scheffer, M},
  journal={Philosophical Transactions of the Royal Society A: Mathematical, Physical and Engineering Sciences},
  volume={370},
  number={1962},
  pages={1185--1204},
  year={2012},
  publisher={The Royal Society Publishing}
}



@article{diks2015critical,
  title={Critical slowing down as an early warning signal for financial crises?},
  author={Diks, Cees and Hommes, Cars and Wang, Juanxi},
  journal={Empirical Economics},
  pages={1--28},
  year={2015},
  publisher={Springer}
}

@article{scheffer2001catastrophic,
  title={Catastrophic shifts in ecosystems},
  author={Scheffer, Marten and Carpenter, Steve and Foley, Jonathan A and Folke, Carl and Walker, Brian},
  journal={Nature},
  volume={413},
  number={6856},
  pages={591},
  year={2001},
  publisher={Nature Publishing Group}
}

@article{scheffer2009early,
  title={Early-warning signals for critical transitions},
  author={Scheffer, Marten and Bascompte, Jordi and Brock, William A and Brovkin, Victor and Carpenter, Stephen R and Dakos, Vasilis and Held, Hermann and Van Nes, Egbert H and Rietkerk, Max and Sugihara, George},
  journal={Nature},
  volume={461},
  number={7260},
  pages={53},
  year={2009},
  publisher={Nature Publishing Group}
}

@article{welch2008importance,
  title={The importance of language in international knowledge transfer},
  author={Welch, Denice E and Welch, Lawrence S},
  journal={Management International Review},
  volume={48},
  number={3},
  pages={339--360},
  year={2008},
  publisher={Springer}
}


@article{bracken2006you,
  title={‘What do you mean?’The importance of language in developing interdisciplinary research},
  author={Bracken, Louise J and Oughton, Elizabeth A},
  journal={Transactions of the Institute of British Geographers},
  volume={31},
  number={3},
  pages={371--382},
  year={2006},
  publisher={Wiley Online Library}
}

@article{welch2005speaking,
  title={Speaking in tongues: The importance of language in international management processes},
  author={Welch, Denice and Welch, Lawrence and Piekkari, Rebecca},
  journal={International Studies of Management \& Organization},
  volume={35},
  number={1},
  pages={10--27},
  year={2005},
  publisher={Taylor \& Francis}
}

@article{patriarca2004modeling,
  title={Modeling language competition},
  author={Patriarca, Marco and Lepp{\"a}nen, Teemu},
  journal={Physica A: Statistical Mechanics and its Applications},
  volume={338},
  number={1-2},
  pages={296--299},
  year={2004},
  publisher={Elsevier}
}

@article{de1993evolving,
  title={The evolving European language system: A theory of communication potential and language competition},
  author={De Swaan, Abram},
  journal={International political science review},
  volume={14},
  number={3},
  pages={241--255},
  year={1993},
  publisher={Sage Publications Sage CA: Thousand Oaks, CA}
}

@article{bates1987competition,
  title={Competition, variation, and language learning},
  author={Bates, Elizabeth and MacWhinney, Brian and MacWhinney, B},
  journal={Mechanisms of language acquisition},
  pages={157--193},
  year={1987}
}

@article{patriarca2004modeling,
  title={Modeling language competition},
  author={Patriarca, Marco and Lepp{\"a}nen, Teemu},
  journal={Physica A: Statistical Mechanics and its Applications},
  volume={338},
  number={1-2},
  pages={296--299},
  year={2004},
  publisher={Elsevier}
}

@article{amato2018dynamics,
  title={The dynamics of norm change in the cultural evolution of language},
  author={Amato, Roberta and Lacasa, Lucas and D{\'\i}az-Guilera, Albert and Baronchelli, Andrea},
  journal={Proceedings of the National Academy of Sciences},
  volume={115},
  number={33},
  pages={8260--8265},
  year={2018},
  publisher={National Acad Sciences}
}

@article{abrams2003linguistics,
  title={Linguistics: Modelling the dynamics of language death},
  author={Abrams, Daniel M and Strogatz, Steven H},
  journal={Nature},
  volume={424},
  number={6951},
  pages={900},
  year={2003},
  publisher={Nature Publishing Group}
}

@article{pinasco2006coexistence,
  title={Coexistence of languages is possible},
  author={Pinasco, Juan Pablo and Romanelli, Liliana},
  journal={Physica A: Statistical Mechanics and its Applications},
  volume={361},
  number={1},
  pages={355--360},
  year={2006},
  publisher={Elsevier}
}

@article{niyi1994linguistic,
  title={Linguistic unification and language rights},
  author={NIYI AKINNASO, F},
  journal={Applied Linguistics},
  volume={15},
  number={2},
  pages={139--168},
  year={1994},
  publisher={Oxford University Press}
}

@book{de2003languages,
  title={When languages collide: Perspectives on language conflict, language competition, and language coexistence},
  author={De Mesquita, Bruce Bueno and DeStefano, Johanna},
  year={2003},
  publisher={Ohio State University Press}
}

@article{cavallaro2014language,
  title={Language in Singapore: From multilingualism to English plus},
  author={Cavallaro, Francesco and Ng, Bee Chin},
  journal={Challenging the monolingual mindset},
  volume={156},
  pages={33},
  year={2014},
  publisher={Multilingual Matters Ltd. Bristol}
}

@article{yun2016possibility,
  title={The possibility of coexistence and co-development in language competition: ecology--society computational model and simulation},
  author={Yun, Jian and Shang, Song-Chao and Wei, Xiao-Dan and Liu, Shuang and Li, Zhi-Jie},
  journal={SpringerPlus},
  volume={5},
  number={1},
  pages={855},
  year={2016},
  publisher={Nature Publishing Group}
}

@book{brenzinger2008language,
  title={Language diversity endangered},
  author={Brenzinger, Matthias},
  volume={181},
  year={2008},
  publisher={Walter de Gruyter}
}

@article{sole2010diversity,
  title={Diversity, competition, extinction: the ecophysics of language change},
  author={Sol{\'e}, Ricard V and Corominas-Murtra, Bernat and Fortuny, Jordi},
  journal={Journal of The Royal Society Interface},
  volume={7},
  number={53},
  pages={1647--1664},
  year={2010},
  publisher={The Royal Society}
}

@article{patriarca2012modeling,
  title={Modeling two-language competition dynamics},
  author={Patriarca, Marco and Castell{\'o}, Xavier and Uriarte, Jos{\'e} Ram{\'o}n and Egu{\'\i}luz, V{\'\i}ctor M and San Miguel, Maxi},
  journal={Advances in Complex Systems},
  volume={15},
  number={03n04},
  pages={1250048},
  year={2012},
  publisher={World Scientific}
}

@article{fujie2013model,
  title={A model of competition among more than two languages},
  author={Fujie, Ryo and Aihara, Kazuyuki and Masuda, Naoki},
  journal={Journal of Statistical Physics},
  volume={151},
  number={1-2},
  pages={289--303},
  year={2013},
  publisher={Springer}
}

@article{patriarca2009influence,
  title={Influence of geography on language competition},
  author={Patriarca, Marco and Heinsalu, Els},
  journal={Physica A: Statistical Mechanics and its Applications},
  volume={388},
  number={2-3},
  pages={174--186},
  year={2009},
  publisher={Elsevier}
}

@article{costa2003another,
  title={Another look at cross-language competition in bilingual speech production: Lexical and phonological factors},
  author={Costa, Albert and Colom{\'e}, {\`A}ngels and G{\'o}mez, Olga and Sebasti{\'a}n-Gall{\'e}s, Nuria},
  journal={Bilingualism: Language and Cognition},
  volume={6},
  number={3},
  pages={167--179},
  year={2003},
  publisher={Cambridge University Press}
}

@techreport{you2018language,
  title={Language Unification, Labor and Ideology},
  author={You, Yang},
  year={2018},
  institution={Working Paper}
}
\begin{CJK*}{GBK}{song}

@article{languages,
  title={ From multi-dialectal society to Cantonese dominant: A survey of language shift in Hong Kong 1949--1971},
  author={Lau, Chun Fat and So, Daniel WC},
  journal={The Journal of Chinese Sociolinguistics},
  pages={89--104},
  year={2005}
}


@Misc{LanguageHongkong,
howpublished = {\url{https://www.bycensus2016.gov.hk/tc/Snapshot-08.html}},
title = {"Use of Language by Hong Kong Population"},
}

\end{CJK*}

@article{bacon1998charting,
  title={Charting multilingualism: Language censuses and language surveys in Hong Kong},
  author={Bacon-Shone, John and Bolton, Kingsley},
  journal={Language in Hong Kong at century’s end},
  pages={43--90},
  year={1998},
  publisher={Hong Kong University Press Hong Kong}
}


@article{leong2014study,
  title={A Study of Attitudes towards the Speak Mandarin Campaign in Singapore.},
  author={Leong, NG Chin},
  journal={Intercultural Communication Studies},
  volume={23},
  number={3},
  year={2014}
}

@misc{suhonen2011speak,
  title={Speak Good English Movement in Singapore: Reactions in Social and Traditional Media},
  author={Suhonen, Lari-Valtteri},
  year={2011}
}

@article{patriarca2012modeling,
  title={Modeling two-language competition dynamics},
  author={Patriarca, Marco and Castell{\'o}, Xavier and Uriarte, Jos{\'e} Ram{\'o}n and Egu{\'\i}luz, V{\'\i}ctor M and San Miguel, Maxi},
  journal={Advances in Complex Systems},
  volume={15},
  number={03n04},
  pages={1250048},
  year={2012},
  publisher={World Scientific}
}

@article{kuhl2011early,
  title={Early language learning and literacy: neuroscience implications for education},
  author={Kuhl, Patricia K},
  journal={Mind, Brain, and Education},
  volume={5},
  number={3},
  pages={128--142},
  year={2011},
  publisher={Wiley Online Library}
}

@article{nishi2013collective,
  title={Collective opinion formation model under Bayesian updating and confirmation bias},
  author={Nishi, Ryosuke and Masuda, Naoki},
  journal={Physical Review E},
  volume={87},
  number={6},
  pages={062123},
  year={2013},
  publisher={APS}
}

@article{pinasco2006coexistence,
  title={Coexistence of languages is possible},
  author={Pinasco, Juan Pablo and Romanelli, Liliana},
  journal={Physica A: Statistical Mechanics and its Applications},
  volume={361},
  number={1},
  pages={355--360},
  year={2006},
  publisher={Elsevier}
}

@article{stauffer2007microscopic,
  title={Microscopic Abrams--Strogatz model of language competition},
  author={Stauffer, Dietrich and Castell{\'o}, Xavier and Eguiluz, Victor M and San Miguel, Maxi},
  journal={Physica A: Statistical Mechanics and its Applications},
  volume={374},
  number={2},
  pages={835--842},
  year={2007},
  publisher={Elsevier}
}

@article{castellano2009statistical,
  title={Statistical physics of social dynamics},
  author={Castellano, Claudio and Fortunato, Santo and Loreto, Vittorio},
  journal={Reviews of modern physics},
  volume={81},
  number={2},
  pages={591},
  year={2009},
  publisher={APS}
}

@book{tomasello2009constructing,
  title={Constructing a language},
  author={Tomasello, Michael},
  year={2009},
  publisher={Harvard university press}
}

@book{de2005second,
  title={Second language acquisition: An advanced resource book},
  author={De Bot, Kees and Lowie, Wander and Verspoor, Marjolyn and Verspoor, Marjolijn Henri{\"e}tte},
  year={2005},
  publisher={Psychology Press}
}

@article{dijkstra1998simulating,
  title={Simulating cross-language competition with the bilingual interactive activation model.},
  author={Dijkstra, Ton and Van Heuven, Walter JB and Grainger, Jonathan},
  journal={Psychologica Belgica},
  year={1998},
  publisher={Belgian Association for Psychological Science}
}

@article{abrams2003linguistics,
  title={Linguistics: Modelling the dynamics of language death},
  author={Abrams, Daniel M and Strogatz, Steven H},
  journal={Nature},
  volume={424},
  number={6951},
  pages={900},
  year={2003},
  publisher={Nature Publishing Group}
}

@article{lieberman2007quantifying,
  title={Quantifying the evolutionary dynamics of language},
  author={Lieberman, Erez and Michel, Jean-Baptiste and Jackson, Joe and Tang, Tina and Nowak, Martin A},
  journal={Nature},
  volume={449},
  number={7163},
  pages={713},
  year={2007},
  publisher={Nature Publishing Group}
}

@article{zhang2013principles,
  title={Principles of parametric estimation in modeling language competition},
  author={Zhang, Menghan and Gong, Tao},
  journal={Proceedings of the National Academy of Sciences},
  volume={110},
  number={24},
  pages={9698--9703},
  year={2013},
  publisher={National Acad Sciences}
}

@article{castellano2009statistical,
  title={Statistical physics of social dynamics},
  author={Castellano, Claudio and Fortunato, Santo and Loreto, Vittorio},
  journal={Reviews of modern physics},
  volume={81},
  number={2},
  pages={591},
  year={2009},
  publisher={APS}
}

@article{loreto2011statistical,
  title={Statistical physics of language dynamics},
  author={Loreto, Vittorio and Baronchelli, Andrea and Mukherjee, Animesh and Puglisi, Andrea and Tria, Francesca},
  journal={Journal of Statistical Mechanics: Theory and Experiment},
  volume={2011},
  number={04},
  pages={P04006},
  year={2011},
  publisher={IOP Publishing}
}

@article{michel2011quantitative,
  title={Quantitative analysis of culture using millions of digitized books},
  author={Michel, Jean-Baptiste and Shen, Yuan Kui and Aiden, Aviva Presser and Veres, Adrian and Gray, Matthew K and Pickett, Joseph P and Hoiberg, Dale and Clancy, Dan and Norvig, Peter and Orwant, Jon and others},
  journal={science},
  volume={331},
  number={6014},
  pages={176--182},
  year={2011},
  publisher={American Association for the Advancement of Science}
}

@article{petersen2012statistical,
  title={Statistical laws governing fluctuations in word use from word birth to word death},
  author={Petersen, Alexander M and Tenenbaum, Joel and Havlin, Shlomo and Stanley, H Eugene},
  journal={Scientific reports},
  volume={2},
  pages={313},
  year={2012},
  publisher={Nature Publishing Group}
}

@book{ginsburgh2011many,
  title={How many languages do we need?: The economics of linguistic diversity},
  author={Ginsburgh, Victor and Weber, Shiomo},
  year={2011},
  publisher={Princeton University Press}
}

\begin{thebibliography}{10}

\bibitem{welch2008importance}
Denice~E Welch and Lawrence~S Welch.
\newblock The importance of language in international knowledge transfer.
\newblock {\em Management International Review}, 48(3):339--360, 2008.

\bibitem{bracken2006you}
Louise~J Bracken and Elizabeth~A Oughton.
\newblock ‘what do you mean?’the importance of language in developing
  interdisciplinary research.
\newblock {\em Transactions of the Institute of British Geographers},
  31(3):371--382, 2006.

\bibitem{welch2005speaking}
Denice Welch, Lawrence Welch, and Rebecca Piekkari.
\newblock Speaking in tongues: The importance of language in international
  management processes.
\newblock {\em International Studies of Management \& Organization},
  35(1):10--27, 2005.

\bibitem{patriarca2004modeling}
Marco Patriarca and Teemu Lepp{\"a}nen.
\newblock Modeling language competition.
\newblock {\em Physica A: Statistical Mechanics and its Applications},
  338(1-2):296--299, 2004.

\bibitem{bates1987competition}
Elizabeth Bates, Brian MacWhinney, and B~MacWhinney.
\newblock Competition, variation, and language learning.
\newblock {\em Mechanisms of language acquisition}, pages 157--193, 1987.

\bibitem{amato2018dynamics}
Roberta Amato, Lucas Lacasa, Albert D{\'\i}az-Guilera, and Andrea Baronchelli.
\newblock The dynamics of norm change in the cultural evolution of language.
\newblock {\em Proceedings of the National Academy of Sciences},
  115(33):8260--8265, 2018.

\bibitem{patriarca2012modeling}
Marco Patriarca, Xavier Castell{\'o}, Jos{\'e}~Ram{\'o}n Uriarte, V{\'\i}ctor~M
  Egu{\'\i}luz, and Maxi San~Miguel.
\newblock Modeling two-language competition dynamics.
\newblock {\em Advances in Complex Systems}, 15(03n04):1250048, 2012.

\bibitem{de2003languages}
Bruce~Bueno De~Mesquita and Johanna DeStefano.
\newblock {\em When languages collide: Perspectives on language conflict,
  language competition, and language coexistence}.
\newblock Ohio State University Press, 2003.

\bibitem{abrams2003linguistics}
Daniel~M Abrams and Steven~H Strogatz.
\newblock Linguistics: Modelling the dynamics of language death.
\newblock {\em Nature}, 424(6951):900, 2003.

\bibitem{costa2003another}
Albert Costa, {\`A}ngels Colom{\'e}, Olga G{\'o}mez, and Nuria
  Sebasti{\'a}n-Gall{\'e}s.
\newblock Another look at cross-language competition in bilingual speech
  production: Lexical and phonological factors.
\newblock {\em Bilingualism: Language and Cognition}, 6(3):167--179, 2003.

\bibitem{suhonen2011speak}
Lari-Valtteri Suhonen.
\newblock Speak good english movement in singapore: Reactions in social and
  traditional media, 2011.

\bibitem{you2018language}
Yang You.
\newblock Language unification, labor and ideology.
\newblock Technical report, Working Paper, 2018.

\bibitem{tomasello2009constructing}
Michael Tomasello.
\newblock {\em Constructing a language}.
\newblock Harvard university press, 2009.

\bibitem{de2005second}
Kees De~Bot, Wander Lowie, Marjolyn Verspoor, and Marjolijn~Henri{\"e}tte
  Verspoor.
\newblock {\em Second language acquisition: An advanced resource book}.
\newblock Psychology Press, 2005.

\bibitem{dijkstra1998simulating}
Ton Dijkstra, Walter~JB Van~Heuven, and Jonathan Grainger.
\newblock Simulating cross-language competition with the bilingual interactive
  activation model.
\newblock {\em Psychologica Belgica}, 1998.

\bibitem{pinasco2006coexistence}
Juan~Pablo Pinasco and Liliana Romanelli.
\newblock Coexistence of languages is possible.
\newblock {\em Physica A: Statistical Mechanics and its Applications},
  361(1):355--360, 2006.

\bibitem{nishi2013collective}
Ryosuke Nishi and Naoki Masuda.
\newblock Collective opinion formation model under bayesian updating and
  confirmation bias.
\newblock {\em Physical Review E}, 87(6):062123, 2013.

\bibitem{michel2011quantitative}
Jean-Baptiste Michel, Yuan~Kui Shen, Aviva~Presser Aiden, Adrian Veres,
  Matthew~K Gray, Joseph~P Pickett, Dale Hoiberg, Dan Clancy, Peter Norvig, Jon
  Orwant, et~al.
\newblock Quantitative analysis of culture using millions of digitized books.
\newblock {\em science}, 331(6014):176--182, 2011.

\bibitem{ginsburgh2011many}
Victor Ginsburgh and Shiomo Weber.
\newblock {\em How many languages do we need?: The economics of linguistic
  diversity}.
\newblock Princeton University Press, 2011.

\bibitem{castellano2009statistical}
Claudio Castellano, Santo Fortunato, and Vittorio Loreto.
\newblock Statistical physics of social dynamics.
\newblock {\em Reviews of modern physics}, 81(2):591, 2009.

\bibitem{stauffer2007microscopic}
Dietrich Stauffer, Xavier Castell{\'o}, Victor~M Eguiluz, and Maxi San~Miguel.
\newblock Microscopic abrams--strogatz model of language competition.
\newblock {\em Physica A: Statistical Mechanics and its Applications},
  374(2):835--842, 2007.

\bibitem{sole2010diversity}
Ricard~V Sol{\'e}, Bernat Corominas-Murtra, and Jordi Fortuny.
\newblock Diversity, competition, extinction: the ecophysics of language
  change.
\newblock {\em Journal of The Royal Society Interface}, 7(53):1647--1664, 2010.

\bibitem{lieberman2007quantifying}
Erez Lieberman, Jean-Baptiste Michel, Joe Jackson, Tina Tang, and Martin~A
  Nowak.
\newblock Quantifying the evolutionary dynamics of language.
\newblock {\em Nature}, 449(7163):713, 2007.

\bibitem{zhang2013principles}
Menghan Zhang and Tao Gong.
\newblock Principles of parametric estimation in modeling language competition.
\newblock {\em Proceedings of the National Academy of Sciences},
  110(24):9698--9703, 2013.

\bibitem{loreto2011statistical}
Vittorio Loreto, Andrea Baronchelli, Animesh Mukherjee, Andrea Puglisi, and
  Francesca Tria.
\newblock Statistical physics of language dynamics.
\newblock {\em Journal of Statistical Mechanics: Theory and Experiment},
  2011(04):P04006, 2011.

\bibitem{petersen2012statistical}
Alexander~M Petersen, Joel Tenenbaum, Shlomo Havlin, and H~Eugene Stanley.
\newblock Statistical laws governing fluctuations in word use from word birth
  to word death.
\newblock {\em Scientific reports}, 2:313, 2012.

\bibitem{fujie2013model}
Ryo Fujie, Kazuyuki Aihara, and Naoki Masuda.
\newblock A model of competition among more than two languages.
\newblock {\em Journal of Statistical Physics}, 151(1-2):289--303, 2013.

\bibitem{cavallaro2014language}
Francesco Cavallaro and Bee~Chin Ng.
\newblock Language in singapore: From multilingualism to english plus.
\newblock {\em Challenging the monolingual mindset}, 156:33, 2014.

\bibitem{languages}
Chun~Fat Lau and Daniel~WC So.
\newblock From multi-dialectal society to cantonese dominant: A survey of
  language shift in hong kong 1949--1971.
\newblock {\em The Journal of Chinese Sociolinguistics}, pages 89--104, 2005.

\bibitem{LanguageHongKong}
"use of language by hong kong population".
\newblock \url{https://www.bycensus2016.gov.hk/tc/Snapshot-08.html}.

\bibitem{bacon1998charting}
John Bacon-Shone and Kingsley Bolton.
\newblock Charting multilingualism: Language censuses and language surveys in
  hong kong.
\newblock {\em Language in Hong Kong at century’s end}, pages 43--90, 1998.

\bibitem{leong2014study}
NG~Chin Leong.
\newblock A study of attitudes towards the speak mandarin campaign in
  singapore.
\newblock {\em Intercultural Communication Studies}, 23(3), 2014.

\bibitem{scheffer2009early}
Marten Scheffer, Jordi Bascompte, William~A Brock, Victor Brovkin, Stephen~R
  Carpenter, Vasilis Dakos, Hermann Held, Egbert~H Van~Nes, Max Rietkerk, and
  George Sugihara.
\newblock Early-warning signals for critical transitions.
\newblock {\em Nature}, 461(7260):53, 2009.

\bibitem{scheffer2001catastrophic}
Marten Scheffer, Steve Carpenter, Jonathan~A Foley, Carl Folke, and Brian
  Walker.
\newblock Catastrophic shifts in ecosystems.
\newblock {\em Nature}, 413(6856):591, 2001.

\bibitem{diks2015critical}
Cees Diks, Cars Hommes, and Juanxi Wang.
\newblock Critical slowing down as an early warning signal for financial
  crises?
\newblock {\em Empirical Economics}, pages 1--28, 2015.

\bibitem{lenton2012early}
TM~Lenton, VN~Livina, V~Dakos, EH~Van~Nes, and M~Scheffer.
\newblock Early warning of climate tipping points from critical slowing down:
  comparing methods to improve robustness.
\newblock {\em Philosophical Transactions of the Royal Society A: Mathematical,
  Physical and Engineering Sciences}, 370(1962):1185--1204, 2012.

\bibitem{patriarca2009influence}
Marco Patriarca and Els Heinsalu.
\newblock Influence of geography on language competition.
\newblock {\em Physica A: Statistical Mechanics and its Applications},
  388(2-3):174--186, 2009.

\bibitem{kuhl2011early}
Patricia~K Kuhl.
\newblock Early language learning and literacy: neuroscience implications for
  education.
\newblock {\em Mind, Brain, and Education}, 5(3):128--142, 2011.

\end{thebibliography}

\end{document}